\documentclass[onecolumn, 10pt, amsmath, amssymb, aps, prd]{revtex4-2}
\usepackage{hyperref}
\usepackage{slashed}
\usepackage{color}
\usepackage{graphicx}
\graphicspath{{figures/}}
\allowdisplaybreaks[1]
\usepackage{multirow}
\usepackage{mathtools}

\begin{document}
\title{Gravitational form factors of pion in a nonlocal quark model}
\author{Vladimir Voronin}
\email{voronin@theor.jinr.ru}
\affiliation{Joint Institute for Nuclear Research, 141980 Dubna, Moscow Region, Russia}
\begin{abstract}
Interaction with external gravitational field is introduced into a nonlocal quark model using the equivalence principle. This allows to define the energy-momentum tensor in flat space and evaluate its hadronic matrix element with external pions. It is found that bubble diagrams give a significant contribution to the corresponding gravitational form factors.

\end{abstract}
\maketitle
\mathtoolsset{showonlyrefs=true,showmanualtags=true}
\section{Introduction}
Gravitational form factors describe the response of non-pointlike hadrons to applied gravitational field. Since the gravity interacts with energy-momentum tensor, this property is contained in the matrix element
\begin{equation}
\langle p',s'|T_{\mu\nu}(x)|p,s\rangle,
\end{equation}
where $T_{\mu\nu}$ is the energy-momentum tensor of quantum chromodynamics (QCD), $p,s$ and $p',s'$ are initial and final momenta and polarizations of a hadron. Gravitational form factors can be related to distributions of mass, momentum, energy, angular momentum, pressure and shear forces inside hadrons~\cite{Polyakov:2002yz}, and thus provide an additional insight into the structure of hadrons. Of particular interest is the so-called $D$-term which is not constrained by Ward identity (see e.g. review~\cite{Polyakov:2018zvc}). Gravitational form factors are unlikely to be measured directly, but they are related to generalized parton distributions~\cite{Ji:1996ek,Ji:1996nm}, so they can be extracted from experimentally observed hard exclusive processes~\cite{Burkert:2018bqq,Duran:2022xag}.

The present paper is focused on pion, and in this case the hadronic matrix element can be parametrized as~\cite{Polyakov:2018zvc}
\begin{equation}
\label{tensor_parametrization}
\langle p'|T_{\mu\nu}(x)|p\rangle=\left[2P_\mu P_\nu A(t)+\frac12\left(\Delta_\mu\Delta_\nu-g_{\mu\nu}\Delta^2\right)D(t)+2M_\pi^2\bar{c}(t)g_{\mu\nu}\right]\exp\left[i(p'-p)x\right].
\end{equation}
Here $P=\frac{1}{2}(p+p'), \Delta=p'-p, t=\Delta^2$.
There is a restriction following from the Ward identity~\cite{Brout:1966oea}
\begin{equation}
A(0)=1,\ \bar{c}(t)=0.
\end{equation}
Additionally, soft pion theorems~\cite{Novikov:1980fa,Voloshin:1980zf,Polyakov:1998ze} predict that the value of $D(0)$, which is commonly called $D$-term, should be equal to $-1$ for massless pions. The gravitational form factors of pion were studied by the methods of lattice QCD in Refs.~\cite{Shanahan:2018pib,Pefkou:2021fni,Hackett:2023nkr}.

The explicit calculation of gravitational form factors in the present paper is performed in the framework of the nonlocal quark model given by the functional
\begin{equation}
Z=N\int \mathcal{D}\bar{q}\mathcal{D}q\mathcal{D}\phi_\pi\exp\left\{\int d^4x\ \sum_f\bar{q}_f(x)\left(i\slashed{\partial}-m_f\right)q_f(x)-\frac12 h^2 M_0^2\phi^2(x)+h\phi(x)\bar{q}(x)V(x)q(x)\right\},
\end{equation}
where $q_f$ and $\phi$ are quark and pion fields. The gluons do not appear explicitly but rather through the nonlocal vertex $V(x)$ which stems from the gluon propagator as explained in Section~\ref{section_description}. In Section~\ref{section_gravity}, the principle of equivalence is used to include interaction with external gravitational field into this nonlocal model. Then one can use the definition of the Hilbert energy-momentum tensor and find the relevant set of Feynman diagrams. These include local interactions with gravity which are essentially of the same type that emerges in the rainbow-ladder approximation in the Bethe--Salpeter approach~\cite{Xu:2023izo}. But there is also an additional significant contribution due to so-called bubble diagrams, which appear when gauge interaction is consistently introduced in nonlocal or bound-state models~\cite{Terning:1991yt,Gross:1987bu,Woloshyn:1975vg,Bentz:1985uji}.
The results are presented in Section~\ref{section_results}.

\section{Description of the model\label{section_description}}
First, consider the model of hadronization in external (anti-)self-dual gluon field~\cite{Efimov:1995uz,Burdanov:1996uw,Kalloniatis:2003sa,Nedelko:2016gdk} which motivates a simplified model mentioned in Introduction. It allows to describe essential phenomena in low-energy QCD such as confinement and chiral symmetry breaking, and evaluate meson observables as well. In particular, confinement is due to the background Abelian (anti-)self-dual gluon field $B$, which provides that the propagators of color-charged fields do not have singularities in finite complex momentum plane~\cite{Leutwyler:1980ev}. Detailed motivation of the vacuum structure and model of hadronization is given in Refs.~\cite{Kalloniatis:2003sa,Nedelko:2016gdk}.

After bosonization of one-gluon exchange in the presence of homogeneous Abelian (anti-)self-dual gluon field $B$ with strength $\Lambda$, one finds the generating functional for mesons and quarks which can be written as~\cite{Efimov:1995uz}
\begin{equation}
\label{generating_functional_with_field}
Z=N \int dB\mathcal{D}\bar{q}\mathcal{D}q\mathcal{D}\phi_\mathcal{Q}\exp\left\{\int d^4x  \bar{q}(x)\left(i\slashed{\partial}+\slashed{B}-m\right)q(x)+\sum_\mathcal{Q}\left[-\frac{\Lambda^2}{2g^2C_J}\phi_\mathcal{Q}^2(x)+\phi_\mathcal{Q}(x)J_\mathcal{Q}(x)\right]\right\}.
\end{equation}
Here integration over $B$ denotes averaging over homogeneous background field configurations, condensed index $Q=aJln$ stands for all quantum numbers of mesons, and anti-Hermitian representation of Dirac matrices is used. 
The current $J_\mathcal{Q}$ stems from decomposition of bilocal quark current $J(x,y)$ over complete set of functions 
in four-dimensional Euclidean space:
\begin{equation}
\label{bilocal_current_decomposition}
J_\mathcal{Q}(x)=\int d^4y \sqrt{D(y)} J_{aJ}\left(x+\xi \frac{y}{\Lambda},x-\xi' \frac{y}{\Lambda}\right)f^{nl}_{\mu_1\dots\mu_l}(y).
\end{equation}
where the weight
\begin{equation}
\sqrt{D(y)}=\sqrt{\frac{1}{4\pi^2y^2}\exp\left(-\frac{y^2}{4}\right)}
\end{equation}
originates from the gluon propagator in homogeneous Abelian (anti-)self-dual gluon field. 
$\xi,\xi'$ provide that $x$ is the center of mass of the bilocal current.
The complete set of functions can be chosen as
\begin{equation}
f^{nl}_{\mu_1\dots\mu_l}(y)=\frac{1}{2}\sqrt{\frac{(l+1)2^l}{\pi^2y^2}}\sqrt{\left(\frac{y^2}{4} \right)^l \exp\left(-\frac{y^2}{4} \right) }L_{nl}\left(\frac{y^2}{4} \right)T^{(l)}_{\mu_1\dots\mu_l}\left(\frac{y}{\sqrt{y^2}}\right),
\end{equation}
where $L_{nk}$ are normalized Laguerre polynomials 
and $T^{(l)}_{\mu_1\dots\mu_l}$ are irreducible tensors of four-dimensional rotations. The current $J_\mathcal{Q}$ is related to the Bethe--Salpeter amplitude in agreement with Ref.~\cite{Goldman:1980ww}.

The field $B$ provides confinement of quarks and allows to perform analytical calculations. However, the field $B$ appears in Eq.~\eqref{generating_functional_with_field} as an external field, hence the energy-momentum tensor is not conserved, and the gravitational form factors would not comply with the conventional Ward identity~\cite{Brout:1966oea}. The latter point is not consequential for the Standard Model hadronic observables, because the effective meson action is translation-invariant~\cite{Efimov:1995uz}, and the electroweak symmetry is respected by the field $B$. The functional~\eqref{generating_functional_with_field} was applied to various phenomena in low-energy physics of mesons: masses of mesons, their electroweak and strong interactions~\cite{Nedelko:2016gdk,Nedelko:2016vpj,Nedelko:2021dsh,Nedelko:2022jso,Voronin:2024ngw}. 

In order to include gravity, the generating functional~\eqref{generating_functional_with_field} has to be modified in such way that Ward identities for meson fields are preserved. The covariant derivatives in Eq.~\eqref{generating_functional_with_field} are transformed into ordinary derivatives, but the shape of nonlocal vertices $V_\mathcal{Q}$ is kept unchanged, as well as the value of the vacuum gluon field strength $\Lambda$. 
In the following, we concentrate on ground-state triplet of pions with $\mathcal{Q}=aP00$. The resulting generating functional with translation-invariant Lagrangian is
\begin{equation}
\label{generating_functional_truncated}
Z=N\int \mathcal{D}\bar{q}\mathcal{D}q\mathcal{D}\phi_\pi\exp\left\{\int d^4x\ \sum_f\bar{q}_f(x)\left(i\slashed{\partial}-m_f\right)q_f(x)-\frac12 h^2 M_0^2\phi^2(x)+h\phi(x)J(x)\right\}.
\end{equation}
The current is given by
\begin{align}
\label{current_flat_definition}
\begin{split}
J^a(x)=&\int d^4y \frac{1}{4\pi^2y^2}\exp\left(-\frac{y^2}{4}\right) \bar{q}\left(x+ \xi\frac{y}{\Lambda}\right)i\gamma_5 M^aq\left(x-\xi'\frac{y}{\Lambda}\right)\\
=&\bar{q}\left(x\right)\left[\int d^4y \frac{1}{4\pi^2y^2}\exp\left(-\frac{y^2}{4}\right) \exp\left(\overleftarrow{\partial}\xi\frac{y}{\Lambda}\right)i\gamma_5 M^a \exp\left(-\overrightarrow{\partial}\xi'\frac{y}{\Lambda}\right)\right] q\left(x\right)=\bar{q}(x)V_{aP00}\left(\overleftrightarrow{\partial}\right)q(x)
\end{split}
\end{align}
where $M^a$ are corresponding flavor matrices.
The nonlocal vertex $V_{aP00}=V_a$ can be written as
\begin{equation}
V_a\left(\overleftrightarrow{\partial}\right)=i\gamma_5M^a\int_0^1 dt \exp\left\{\left(\frac{\overleftrightarrow{\partial}}{\Lambda}\right)^2t\right\},\quad \overleftrightarrow{\partial}=\xi\overleftarrow{\partial}-\xi'\overrightarrow{\partial}.
\end{equation}
At large momentum the vertex behaves as $1/p^2$, which makes all diagrams that appear in the present paper finite.

The quark propagators $S(x)$ in the simplified model~\eqref{generating_functional_truncated} are free Dirac ones
\begin{equation}
\left(i\slashed{\partial}-m\right)S(x)=-\delta^{(4)}(x-y).
\end{equation}
In other words, with the removal of the background field $B$, the confinement is lost,
but this problem does not affect the gravitational form factors because the quark production thresholds do not emerge in physically allowed kinematic regime. Moreover, unlike the model given by Eq.~\eqref{generating_functional_with_field}, the pion mass $M$ becomes a phenomenological parameter of the model~\eqref{generating_functional_truncated}. The term $\frac12 h^2M_0^2\phi_\pi^2$ in the Lagrangian provides that the zero of two-point meson correlation function 
\begin{equation}
\Pi(p^2)=h^2M_0^2+h^2\left.\int d^4x \exp(ipx)\ \mathrm{Tr}V\left(\overleftrightarrow{\partial}_x\right)S(x-y)V\left(\overleftrightarrow{\partial}_y\right)S(y-x)\right|_{y=0}
\end{equation}
is located at the physical pion mass in momentum space:
\begin{equation}
\label{mass_eq}
\Pi(-M_\pi^2)=0. 
\end{equation}
To provide correct residue of the meson propagator, the pion field is rescaled as $\phi_\pi \to h\phi_\pi$ 
such that
\begin{equation}
\label{meson-quark_constant}
1=\left.\frac{d}{dp^2}\Pi(p^2)\right|_{p^2=-M^2}.
\end{equation}

\begin{figure}
\includegraphics[width=4cm]{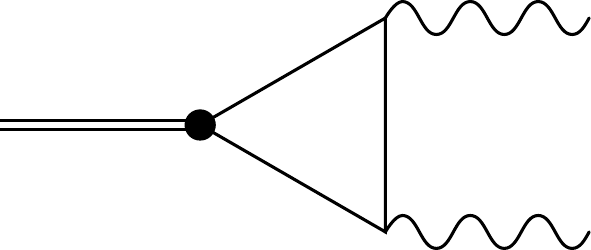}
\caption{\label{figure_pgg}
One-loop diagram contributing to $\pi\to \gamma\gamma$. The second diagram with permuted photon legs is not shown.
}
\end{figure}

The parameters of the model described by Eq.~\eqref{generating_functional_truncated} are given in Table~\ref{table_parameters}. We employ the approximation $m_u=m_d$, so $\xi=\xi'=1/2$.
\begin{table}
\begin{tabular}{c@{\hspace{1em}}c@{\hspace{1em}}c}
\hline\hline
$\Lambda$, MeV&$m_{u/d}$, MeV&$M_0$, MeV\\
\hline
439.9&238.2&123.5\\
\hline\hline
\end{tabular}
\caption{\label{table_parameters}
Parameters of the model~\eqref{generating_functional_truncated} extracted from mass of pion and width of decay $\pi\to\gamma\gamma$.
}
\end{table}
Parameter $\Lambda$ in Table~\ref{table_parameters} is determined in Ref.~\cite{Nedelko:2016gdk} using the model given by Eq.~\eqref{generating_functional_with_field}.
The quark mass in Table~\ref{table_parameters} is such that the decay width of pion into a couple of photons $\Gamma(\pi\to\gamma\gamma)\approx 7.7$ eV~\cite{ParticleDataGroup:2024cfk}, where the amplitude is found from the diagram in Fig.~\ref{figure_pgg}
\begin{equation}
e^2\frac{1}{\sqrt{2}}\left[\left(\frac{2}{3}\right)^2-\left(\frac{1}{3}\right)^2\right]
(-1)\ h\ \mathrm{Tr}V\left(\overleftrightarrow{\partial}_x\right)S(x-y)i\gamma_\mu S(y-z)i\gamma_\nu S(z-x)
\end{equation}
and its crossing. Diagrams with nonlocal vertices are most conveniently evaluated with the method outlined in Appendix~\ref{appendix_diagrams_expressions}.
One finds from formula~\eqref{meson-quark_constant} that $h=5.5$ with the parameters given in Table~\ref{table_parameters}.

The model defined by Eq.~\eqref{generating_functional_truncated} is a type of models with separable interaction considered in Ref.~\cite{Anikin:1995cf}, but the form of the vertex $V_{aP00}$ is determined by the gluon propagator in Eq.~\eqref{bilocal_current_decomposition}.

\section{Interaction with gravity\label{section_gravity}}
The gluon field appears in the energy-momentum tensor of QCD but is integrated out in Eq.~\eqref{generating_functional_truncated}. It is possible to find combined contribution to the gravitational form factors from both gluons and quarks if one introduces gravitational field in Eq.~\eqref{generating_functional_truncated}. For this purpose it is sufficient to introduce external torsionless gravitational field in four-dimensional Euclidean space using the equivalence principle, which can be achieved with the minimal substitution
\begin{gather}
\overrightarrow{\partial}_a\to \overrightarrow{D}_a=e_a^\alpha \overrightarrow{D}_\alpha=e_a^\alpha\left(\overrightarrow{\partial}_\alpha-\Omega_\alpha\right),\\
\overleftarrow{\partial}_a\to \overleftarrow{D}_a=e_a^\alpha \overleftarrow{D}_\alpha=e_a^\alpha\left(\overleftarrow{\partial}_\alpha+\Omega_\alpha\right),\\
d^4x\to d^4x\ \left|\det e_a^\alpha\right|=d^4x\ e(x)=d^4x\ \sqrt{\det|g_{\mu\nu}|},\\
\label{connection}
\Omega_\alpha=\frac{1}{8}\omega_{\alpha bc}\left(\gamma^b\gamma^c-\gamma^c\gamma^b\right).
\end{gather}
The frame field (tetrad, vierbein) $e^a_\alpha(x)$ and its inverse are defined by
\begin{equation}
g_{\mu\nu}=e^a_\mu \eta_{ab}e^b_\nu,\quad e^a_\alpha e_b^\alpha=e^{a\alpha} e_{b\alpha}=\delta_b^a,\quad e_{a\alpha}=\eta_{ab}e^b_\alpha=g_{\alpha\beta}e_a^\beta,
\end{equation}
where $g_{\mu\nu}$ is metric in curved space and $\eta_{ab}=\delta_{ab}$ is flat metric.
The spin connection is given by
\begin{equation}
\label{spin_connection}
\omega_{\mu ab}=\frac{1}{2}e_a^\nu (\partial_\mu e_{b\nu}-\partial_\nu e_{b\mu})-
\frac{1}{2}e_b^\nu (\partial_\mu e_{a\nu}-\partial_\nu e_{a\mu})-
\frac{1}{2}e_a^\rho e_b^\sigma e_\mu^c (\partial_\rho e_{c\sigma}-\partial_\sigma e_{c\rho}).
\end{equation}
The generating functional becomes
\begin{gather}
Z=N\int \mathcal{D}\bar{q}\mathcal{D}q\mathcal{D}\phi_\pi\exp\left\{\int d^4x\ e\ \bar{q}\left(i\overrightarrow{\slashed{D}}-m\right)q-\frac12 h^2 M_0^2\phi^2+h\phi J_{g}\right\},\\
\label{nonlocal_current_g}
J_{g}^a(x)=\bar{q}\left(x\right)\left[\int d^4y \frac{1}{4\pi^2y^2}\exp\left(-\frac{y^2}{4}\right) \exp\left(\overleftarrow{D}\xi\frac{y}{\Lambda}\right)i\gamma_5 M^a \exp\left(-\overrightarrow{D}\xi'\frac{y}{\Lambda}\right)\right] q\left(x\right)=\bar{q}\left(x\right)V_g^a(x) q\left(x\right).
\end{gather}
Note that the coordinate $y$ in the current $J_{g}^a$ remains flat. Upon integration over quarks, one finds the effective meson action
\begin{equation}
\label{action_gravity}
S_\text{eff}=-\int d^4x\ e(x)\ \frac12 h^2 M_0^2\phi^2+\mathrm{Tr}\log\left\{1-he(x)\phi(x)V(x)S_g(x,y)\right\},
\end{equation}
where the quark propagator $S_g$ satisfies
\begin{equation}
e(x) \left(i\overrightarrow{\slashed{D}}-m\right)S_g(x,y)=-\delta^{(4)}(x-y).
\end{equation}
Now one can apply the definition
\begin{equation}
\label{energy-momentum_definition}
T_{\mu\nu}=\frac{1}{e}\frac{\delta S_\text{eff}}{\delta e^{a\mu}}e_\nu^a
\end{equation}
which results in the symmetric energy-momentum tensor due to local Lorentz invariance of the action~\eqref{action_gravity}, and evaluate the matrix element of the energy-momentum tensor in flat space.
In the one-loop approximation, one finds
\begin{gather}
\label{matrix_element}
\langle p'|T_{\mu\nu}(z)|p\rangle=\left.\frac{e_\nu^a}{e}\frac{\delta }{\delta e^{a\mu}(z)}\int d^4x \exp(-ipx) \int d^4x' \exp(ip'x')(-1)\left[\delta^{(4)}(x-x')e(x)h^2M_0^2+e(x)e(x')\Pi_g(x,x')\right]\right|_\text{flat},\\
\Pi_g(x,x')=h^2\mathrm{Tr}V_g(x)S_g(x,x')V_g(x')S_g(x',x).
\end{gather}
Variational derivatives can be found using expansion of the frame field around flat space
\begin{equation}
\label{tetrad_expansion}
e_\mu^a\approx\delta_\mu^a+\delta e_\mu^a,\ e^\mu_a\approx\delta^\mu_a+\delta e^\mu_a,\ e(x)\approx 1+\delta_a^\mu\delta e^a_\mu = 1-\delta^a_\mu\delta e_a^\mu.
\end{equation}
The expansions of propagators and vertices are given by (see Appendix~\ref{appendix_expansions})
\begin{gather}
\label{propagator_expansion}
\begin{split}
&S_g(x_1,x_2)=S(x_1-x_2)+\int d^4x\delta e^{a\mu}(x)\delta_a^\nu\\
&\times S(x_1-x)\left(\frac{i}{2}\gamma_\nu\overrightarrow{\partial}_\mu-\frac{i}{2}\overleftarrow{\partial}_\mu\gamma_\nu-\delta_{\mu\nu}\left(\frac{i}{2}\overrightarrow{\slashed{\partial}}-\frac{i}{2}\overleftarrow{\slashed{\partial}}-m\right)+\frac{i}{16}\left\{\overleftarrow{\slashed{\partial}}+ \overrightarrow{\slashed{\partial}},\left[\gamma_\mu,\gamma_\nu\right]\right\}\right)S(x-x_2)+\dots,
\end{split}\\
\label{vertex_g_expansion}
\begin{split}
&\bar{q}(x)V_g(x)q(x)=\bar{q}(x)V\left(\overleftrightarrow{\partial}\right)q(x)+\int d^4z \delta e^{a\mu}(z)\delta_a^\nu\int\frac{d^4p}{(2\pi)^4}\exp (ipz-ipx)\int_0^1d\tau\frac{1}{i\tau}\\
&\times \bar{q}(x)\left\{\left(-\overleftarrow{\partial}_\mu\frac{\partial}{\partial p^\nu}+
\frac{i}{8}\left[\gamma_\mu,\gamma_\nu\right]p^\alpha\frac{\partial}{\partial p^\alpha}+
\frac{i}{8}\left[\gamma_\alpha,\gamma_\mu\right]p^\alpha\frac{\partial}{\partial p^\nu}+
\frac{i}{8}\left[\gamma_\alpha,\gamma_\nu\right]p^\alpha\frac{\partial}{\partial p^\mu}
\right)V(\overleftrightarrow{\partial}-ip\tau\xi)\right.\\
&+\left.\left(-\overrightarrow{\partial}_\mu\frac{\partial}{\partial p^\nu}-
\frac{i}{8}\left[\gamma_\mu,\gamma_\nu\right]p^\alpha\frac{\partial}{\partial p^\alpha}-
\frac{i}{8}\left[\gamma_\alpha,\gamma_\mu\right]p^\alpha\frac{\partial}{\partial p^\nu}-
\frac{i}{8}\left[\gamma_\alpha,\gamma_\nu\right]p^\alpha\frac{\partial}{\partial p^\mu}
\right)V(\overleftrightarrow{\partial}+ip\tau\xi')\right\}q(x)+\dots\\
\end{split}
\end{gather}
where omitted terms are of higher order in $\delta e_a^\mu$.
The one-loop contributions to the matrix element of energy-momentum tensor contained in Eq.~\eqref{matrix_element} can be arranged in the diagrams shown in Fig.~\ref{figure_ppg} for which the explicit formulas are given in Appendix~\ref{appendix_diagrams_expressions}. 
\begin{figure}
\begin{tabular}{c@{\hspace*{1em}}c@{\hspace*{1em}}c@{\hspace*{1em}}c@{\hspace*{1em}}c}
\includegraphics[scale=0.35]{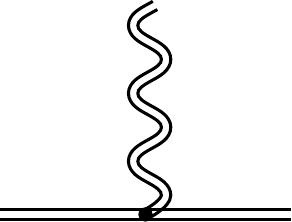}&
\includegraphics[scale=0.35]{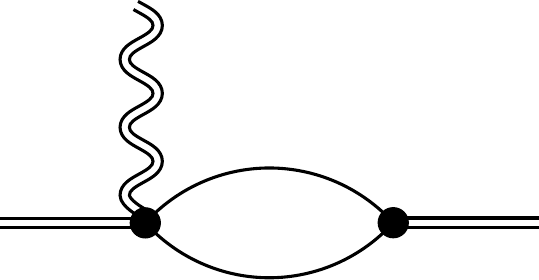}&
\includegraphics[scale=0.35]{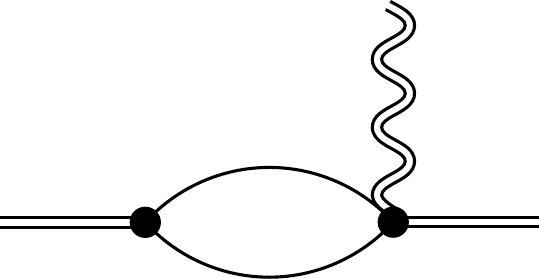}&
\includegraphics[scale=0.35]{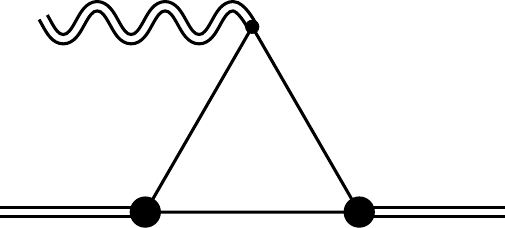}&
\includegraphics[scale=0.35]{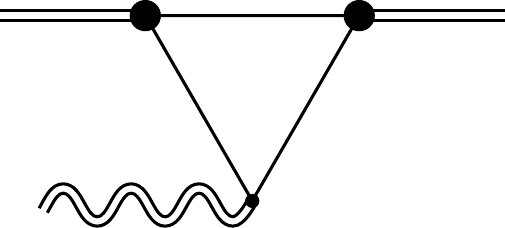}\\
(a)&(b)&(c)&(d)&(e)
\end{tabular}
\caption{\label{figure_ppg}
One-loop diagrams contributing to the matrix element of the energy-momentum tensor. The diagrams related by crossings are not shown. Diagram (a) stems from decomposition of the volume element in powers of $\delta e_a^\mu$, diagrams (b) and (c) originate from decomposition of nonlocal vertices, and diagrams (d) and (e) are generated by decomposition of the quark propagator.}
\end{figure}
In flat Minkowski space, the matrix element of the energy-momentum tensor can be parametrized by formula~\eqref{tensor_parametrization}.
The diagram (a) in Fig.~\ref{figure_ppg} contributes only to $\bar{c}(t)$ and is important for energy-momentum conservation. The contribution due to nonlocal vertices is contained in diagrams (b) and (c). The diagrams (d) and (e) correspond to the impulse approximation of the energy-momentum tensor matrix element in the Bethe--Salpeter approach.

\section{Results and conclusion\label{section_results}}

Nonlocal model~\eqref{generating_functional_truncated} is motivated by model~\eqref{generating_functional_with_field} and retains the shape of its nonlocal vertices which follows from the gluon propagator. These nonlocal vertices encode an important contribution of gluons to the matrix element of the energy-momentum tensor, even though the gluons do not appear explicitly. 

It is known that Noether energy-momentum tensor is not necessarily symmetric or gauge-invariant in QCD. However, one can modify it by adding a total derivative which yields Belinfante-improved tensor (see e.g. discussion in Ref.~\cite{Leader:2013jra}). Definition~\eqref{energy-momentum_definition} automatically yields symmetric conserved energy-momentum tensor that interacts with torsionless gravity, and also allows to extract implicit contribution of gluons from nonlocal vertices of model~\eqref{generating_functional_truncated} in addition to the impulse approximation.

\begin{figure}
\includegraphics[width=3.4in]{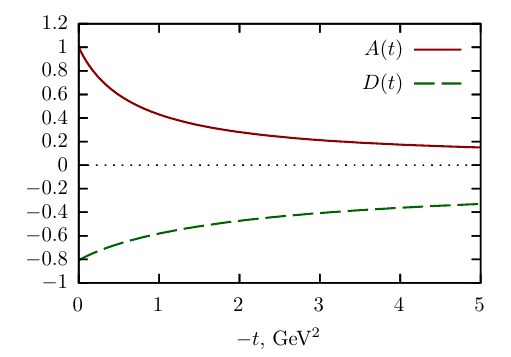}
\caption{\label{figure_gravitational_form_factors}
Gravitational form factors of pion found from diagrams in Fig.~\ref{figure_ppg}.
}
\end{figure}

Upon evaluation of diagrams in Fig.~\ref{figure_ppg} using technique described in Appendix~\ref{appendix_diagrams_expressions} and continuation to Minkowski space, one finds the gravitational form factors that are shown in Fig.~\ref{figure_gravitational_form_factors}.
The value of $D$-term $D(0)$ is roughly $-1$ as predicted by soft pion theorems~\cite{Novikov:1980fa,Voloshin:1980zf,Polyakov:1998ze}.

In the Breit frame, which is specified by the relation $\vec{P}=0$, the gravitational form factors can be used to define the distribution~\cite{Polyakov:2002yz}
\begin{equation}
T_{\mu\nu}(\vec{r})=\int\frac{d^3\Delta}{(2\pi)^3 2E}\exp\left(i\vec{r}\vec{\Delta}\right)\left. \langle p'|T_{\mu\nu}(0)|p\rangle\right|_{\vec{P}=0}
\end{equation}
in the same way as for electromagnetic form factors.
Here $E=\sqrt{M^2+\vec{\Delta}^2/4}$.
Corresponding mean square radius
\begin{equation}
\langle r^2\rangle=\frac{\int d^3r\ r^2 T_{00}(\vec{r})}{\int d^3r\ T_{00}(\vec{r})}=6\left.\frac{dA(t)}{dt}\right|_{t=0}-\frac{3}{4M_\pi^2}(A(0)+2D(0)),
\end{equation}
where $t=-\vec{\Delta}^2$,
evaluates to $\sqrt{\langle r^2\rangle}=1.12\ \text{fm}$ with the form factors shown in Fig.~\ref{figure_gravitational_form_factors}. Comparing this value to the charge radius of pion $0.659\pm 0.004\text{fm}$~\cite{ParticleDataGroup:2024cfk}, one can conclude that the charge in pion is distributed more compactly than mass, although the interpretation of $T_{\mu\nu}(\vec{r})$ as an actual spacial distribution is not exact~\cite{Miller:2018ybm}.

An important observation is that diagrams (b) and (c) in Fig.~\ref{figure_ppg} give a substantial contribution to the gravitational form factors. For example, they are responsible for approximately 8\% of the numerical value of $A(0)$ and 69\% of $D(0)$. Their separate contribution, however, is not a well-defined quantity because only the sum of all diagrams in Fig.~\ref{figure_gravitational_form_factors} yields conserving energy-momentum tensor. In contrast, analogous contributions to the electroweak interactions of hadrons are of minor importance~\cite{Voronin:2024ngw,Faessler:2008ix}. This is not surprising because gravity directly couples to gluons, while the impulse approximation only accounts for interaction with the quark current.

\section{Acknowledgements}
The author is grateful to S. Nedelko for stimulating and beneficial discussions.

\appendix
\section{Expansions of propagators and vertices around flat space\label{appendix_expansions}}
\subsection{Propagator}
The propagator in curved space satisfies equation
\begin{equation}
e(x) \left[ie_a^\mu\gamma^a\left(\partial_\mu-\Omega_\mu\right)-m\right]S_g(x,y)=-\delta^{(4)}(x-y).
\end{equation}
One can transform the propagator as
\begin{align}
S_g=&-\left\{e(x) \left[ie_a^\mu\gamma^a\left(\partial_\mu-\Omega_\mu\right)-m\right]\right\}^{-1}\\
=&-\left(i\slashed{\partial}-m\right)^{-1}\left\{\left(e(x) \left[ie_a^\mu\gamma^a\left(\partial_\mu-\Omega_\mu\right)-m\right]-\left[i\slashed{\partial}-m\right]\right)\left(i\slashed{\partial}-m\right)^{-1}+1\right\}^{-1}\\
=&-\left(i\slashed{\partial}-m\right)^{-1}+\left(i\slashed{\partial}-m\right)^{-1}\left\{e(x) \left[ie_a^\mu\gamma^a\left(\partial_\mu-\Omega_\mu\right)-m\right]-\left[i\slashed{\partial}-m\right]\right\}\left(i\slashed{\partial}-m\right)^{-1}+\dots
\end{align}
Performing integration by parts and changing order of $\Omega$ and $\gamma$ using commutation relations, one finds
\begin{multline}
S_g(x_1,x_2)=
S(x_1-x_2)+\int d^4xS(x_1-x)\left\{\frac{i}{2}e(x)e_a^\mu\gamma^a\left(\overrightarrow{\partial}_\mu-\Omega_\mu\right) -\frac{i}{2}\left(\overleftarrow{\partial}_\mu+\Omega_\mu\right)e(x)e_a^\mu\gamma^a\right.\\
\left.-e(x)m-\left(i\overrightarrow{\slashed{\partial}}-m\right)\right\}S(x-x_2)+\dots
\end{multline}
Expanding the frame field using formula~\eqref{tetrad_expansion}, one finds
\begin{multline}
S_g(x_1,x_2)=
S(x_1-x_2)+\int d^4xS(x_1-x)\left\{\frac{i}{2}\delta e_a^\mu\gamma^a\overrightarrow{\partial}_\mu-\frac{i}{2}\overleftarrow{\partial}_\mu\gamma^a\delta e_a^\mu-\delta_\mu^a \left(\frac{i}{2}\delta e_a^\mu\overrightarrow{\slashed{\partial}}-\frac{i}{2}\overleftarrow{\slashed{\partial}}\delta e_a^\mu-m\delta e_a^\mu\right)\right.\\
\left.-\frac{i}{2}\delta_a^\mu\left(\gamma_a\delta\Omega_\mu+\delta\Omega_\mu\gamma_a\right)\right\}S(x-x_2)+\dots
\end{multline}
Using Eqs.~\eqref{connection},~\eqref{spin_connection} and integrating by parts once again, one obtains Eq.~\eqref{propagator_expansion}.

\subsection{Nonlocal vertex}
Consider the operator
\begin{equation}
\exp\left(\overleftarrow{D}\xi\frac{y}{\Lambda}\right)i\gamma_5 M^a \exp\left(-\overrightarrow{D}\xi'\frac{y}{\Lambda}\right)
\end{equation}
that appears in Eq.~\eqref{nonlocal_current_g}. The exponential of an operator $A$ can be defined as
\begin{equation}
\exp A=\lim_{n\to\infty}\prod_{i=1}^n \exp \frac{A}{n}.
\end{equation}
Consequently, the expansion in small $\delta A$ can be written in the form
\begin{equation}
\label{exp_expansion}
\exp \left(A+\delta A\right)=\exp A+\lim_{n\to\infty}\sum_{j=1}^n \exp \frac{jA}{n}\ \delta A\ \exp\frac{(n-j)A}{n}+\dots
\end{equation}
Using formula~\eqref{exp_expansion}, properties of the shift operator $\exp a\partial$ and taking the limit, one finds
\begin{align}
\exp\left(\overleftarrow{D}\xi\frac{y}{\Lambda}\right)=&\exp\left(\overleftarrow{\partial}\xi\frac{y}{\Lambda}\right)+\int_0^1d\tau\xi y^a\exp\left(\overleftarrow{\partial}\xi\frac{y}{\Lambda}\right)\left[\overleftarrow{\partial}_\mu \delta e_a^\mu\left(x+\tau\xi \frac{y}{\Lambda}\right)+ \delta_a^\mu\delta\Omega_\mu\left(x+\tau\xi \frac{y}{\Lambda}\right)\right]+\dots,\\
\exp\left(-\overrightarrow{D}\xi'\frac{y}{\Lambda}\right)=&\exp\left(-\overrightarrow{\partial}\xi'\frac{y}{\Lambda}\right)-\int_0^1d\tau \xi' y^a\left[\delta e_a^\mu\left(x-\tau\xi' \frac{y}{\Lambda}\right)\overrightarrow{\partial}_\mu- \delta_a^\mu\delta\Omega_\mu\left(x-\tau\xi' \frac{y}{\Lambda}\right)\right]\exp\left(-\overrightarrow{\partial}\xi\frac{y}{\Lambda}\right)+\dots
\end{align}
Now, using formula~\eqref{spin_connection} to expand $\delta\Omega_\mu$, Fourier transformation of $\delta e_a^\mu$ and definition of the vertex $V$ in Eq.~\eqref{current_flat_definition}, one can write the expansion of $V_g$ given by formula~\eqref{vertex_g_expansion}.

\section{Expressions for diagrams\label{appendix_diagrams_expressions}}
It is convenient to introduce shorthand notations for Eqs.~\eqref{propagator_expansion} and~\eqref{vertex_g_expansion}:
\begin{gather}
S_g(x_1,x_2)=S(x_1-x_2)+\int d^4x\delta e^{a\mu}(x)\delta_a^\nu S(x_1-x)\mathcal{O}_{\mu\nu}(x) S(x-x_2)+\dots,
\\
\bar{q}(x)V_g(x)q(x)=\bar{q}(x)V^{(0)}\left(x\right)q(x)+\int d^4z \delta e^{a\mu}(z)\delta_a^\nu \bar{q}(x)V_{\mu\nu}^{(1)}\left(x-z\right)q(x)\dots
\end{gather}
Then contributions of the diagrams in Fig.~\ref{figure_ppg} to the matrix element~\eqref{matrix_element} can be written as
\begin{equation}
\langle p'|T_{\mu\nu}(z)|p\rangle=(\text{a})+(\text{b})+(\text{c})+(\text{d})+(\text{e}),
\end{equation}
where the summands are given by
\begin{gather}
\tag{a}
\delta_{\mu\nu}\exp\left[i(p'-p)z\right]h^2M_0,\\
\tag{b}
\begin{split}
\int d^4x\exp(-ipx)\int d^4x\exp(ip'x')(-1)h^2&\left\{-\delta_{\mu\nu}\delta^{(4)}(x-z)\mathrm{Tr}V^{(0)}(x)S(x-x')V^{(0)}(x')S(x'-x)\right.\\
&\hphantom{\{}\left.+\mathrm{Tr}V_{\mu\nu}^{(1)}(x-z)S(x-x')V^{(0)}(x')S(x'-x)\right\},
\end{split}\\
\tag{c}
\begin{split}
\int d^4x\exp(-ipx)\int d^4x\exp(ip'x')(-1)h^2&\left\{-\delta_{\mu\nu}\delta^{(4)}(x'-z)\mathrm{Tr}V^{(0)}(x)S(x-x')V^{(0)}(x')S(x'-x)\right.\\
&\hphantom{\{}\left.+\mathrm{Tr}V^{(0)}(x)S(x-x')V_{\mu\nu}^{(1)}(x'-z)S(x'-x)\right\},
\end{split}\\
\tag{d}
\begin{split}
\int d^4x\exp(-ipx)\int d^4x\exp(ip'x')(-1)h^2\mathrm{Tr}V^{(0)}(x)S(x-x')V^{(0)}(x')S(x'-z)\mathcal{O}_{\mu\nu}(z)S(z-x),
\end{split}\\
\tag{e}
\begin{split}
\int d^4x\exp(-ipx)\int d^4x\exp(ip'x')(-1)h^2\mathrm{Tr}V^{(0)}(x)S(x-z)\mathcal{O}_{\mu\nu}(z)S(z-x')V^{(0)}(x')S(x'-x).
\end{split}
\end{gather}
The terms proportional to $\delta_{\mu\nu}$ originate from the expansion of $e(x)$.

The explicit expressions are most conveniently obtained using Schwinger representation of propagators
\begin{equation}
S(x-y)= \int_0^\infty ds \int \frac{d^4p}{(2\pi)^4} \exp\left[  - (p^2+m^2) s - ip(x-y)\right]\\
\left( p_\mu\gamma_\mu  + m \right).
\end{equation}
The traces over flavor color are trivial in one-loop case, and the trace over products of Dirac matrices can be evaluated in a computer algebra system such as FORM~\cite{Ruijl:2017dtg}.
The result of action of vertex operators $V$ can be found using
\begin{equation*}
\exp(ipx)\int_0^1 dt \exp\left\{\left(\frac{\overleftrightarrow{\partial}}{\Lambda}\right)^2t\right\}\exp(-iqx)=\exp(ipx)\int_0^1 dt \exp\left\{-\frac{(\xi p+\xi'q)^2}{\Lambda^2}t\right\}\exp(-iqx)
\end{equation*}
which can be deduced from Eq.~\eqref{current_flat_definition}. The momentum integrals become Gaussian and can be evaluated analytically. Then the remaining integrals over proper times originating from the propagators and vertices can be evaluated numerically.
\bibliography{references}
\end{document}